\documentclass[twocolumn]{jpsj2}

\title{\textbf{STM/STS Study on 4a$\times $4a Electronic Charge Order of 
Superconducting Bi$_{2}$Sr$_{2}$CaCu$_{2}$O$_{8+ \delta }$}}

\author{Naoki \textsc{Momono}, Akihiro \textsc{Hashimoto}, Yasuo \textsc{Kobatake}, Migaku \textsc{Oda}, and Masayuki \textsc{Ido}}

\inst{Division of Physics, Graduate School of Science, Hokkaido University, Sapporo 0600810}

\abst{We performed low-bias STM measurements on underdoped Bi2212 crystals, and 
confirmed that a two-dimensional (2D) superstructure with a periodicity of 
four lattice constants (4a) is formed within the Cu-O plane at $T<T_{c}$. 
This 4a$\times $4a superstructure, oriented along the Cu-O bonding 
direction, is nondispersive and more intense in lightly doped samples with a 
zero temperature pseudogap (ZTPG) than in samples with a d-wave gap. The 
nondispersive 4a$\times $4a superstructure was clearly observed within the 
ZTPG or d-wave gap, while it tended to fade out outside the gaps. The 
present results provide a useful test for various models proposed for an 
electronic order hidden in the underdoped region of high-$T_{c}$ cuprates. 
}

\kword{STM/STS, pseudogap, electronic charge ordering, superstructure, cuprate}

\begin{document}
\maketitle

Recently STM/STS studies on the pseudogap state of 
Bi$_{2}$Sr$_{2}$CaCu$_{2}$O$_{8 + \delta }$(Bi2212) at $T>T_{c}$ have 
revealed a nondispersive two-dimensional (2D) superstructure with a 
periodicity of about four lattice constants ($\sim $4a$\times $4a 
superstructure) in the map of energy-resolved differential tunneling 
conductance $dI/dV$, proportional to the local density of states (LDOS).$^{1)}$ 
The nondispersive 4a$\times $4a structure, electronic in origin, was also 
reported in the LDOS maps taken on the zero temperature pseudogap (ZTPG) 
regions of lightly doped Ca$_{2 - x}$Na$_{x}$CuO$_{2}$Cl$_{2}$ (Na-CCOC) and 
Bi2212 samples.$^{2,3)}$ A similar spatial structure was first observed 
around the vortex cores of Bi2212 exhibiting a pseudogap-like V-shaped 
spectrum with no coherence peak.$^{4)}$ Such 2D spatial structures have 
attracted much attention because they can be a possible electronic order 
hidden in the pseudogap state. 

From the LDOS maps taken for the superconducting (SC) state of Bi2212, 
Hoffman et al. and McElroy et al. reported a strongly dispersive 2D 
superstructure.$^{5,6)}$ This superstructure has been successfully explained 
in terms of a quasiparticle scattering interference. Furthermore, Howald et 
al. found a nondispersive $\sim $4a$\times $4a superstructure with 
anisotropy in addition to the weakly dispersive one, and claimed that the 
nondispersive $\sim $4a$\times $4a superstructure will be due to the 
so-called stripe order and coexist with superconductivity.$^{7 - 10)}$ 
However, the nondispersive $\sim $4a$\times $4a superstructure was not 
confirmed in later LDOS measurements on Bi2212 at $T<T_{c}$.$^{1, 5)}$ This 
inconsistency mainly originates from the difficulty in distinguishing the 
nondispersive $\sim $4a$\times $4a superstructure from the weakly dispersive 
$\sim $4a$\times $4a one resulting from quasiparticle scattering 
interference. Since the dispersive features in the LDOS are reduced by 
integrating the LDOS over a wide range of energy, it is desirable to 
investigate the nondispersive features of the LDOS by using a conventional 
STM technique which actually maps the LDOS integrated between $E_{F}$ and 
$E_{F}+eV_{s}$, where $V_{s}$ is the sample bias voltage. To clarify 
whether the nondispersive 4a$\times $4a electronic superstructure persists 
in the SC state ($T<T_{c}$) will promote our understanding of the physics in 
the underdoped region

In the present study, we performed low-bias STM measurements on underdoped 
Bi2212 samples, and succeeded in observing an energy-independent 4a$\times 
$4a electronic superstructure with a substructure having a periodicity of 
4a/3 in the low-bias STM images, which is essentially the same as the 
electronic checkerboard order reported by Howald et al. and Hanaguri et al. 
in LDOS maps of lightly-doped Bi2212 and Na-CCOC.$^{2, 7)}$ The present 4a$\times $4a 
superstructure is accompanied by a d-wave-like gap or ZTPG, and is discussed 
in terms of a wide variety of unusual electronic orders proposed for the 
underdoped region of high-$T_{c}$ cuprates. 

The single crystal of Bi2212 in the present study was grown by the traveling 
solvent floating zone method. We estimated doping level $p$ of the sample from 
the SC critical temperature $T_{c}$ determined from the superconducting 
diamagnetism and the characteristic temperature $T_{max}$ of the 
normal-state magnetic susceptibility; $T_{c}$ and $T_{max}$ follow empirical 
functions of $p$ respectively.$^{11, 12)}$ We performed STM/STS experiments at 
$T\sim $9 K on two samples A and B cut from the same single crystal with 
$T_{c}\sim $72 K ($p\sim $0.11). In the present STM/STS experiments, the 
sample was cleaved in an ultra-high vacuum at $T\sim $9 K just before the 
approach of the STM tip to the cleaved surface in situ. As is well known, 
Bi2212 crystals are usually cleaved between neighboring Bi-O layers. The 
excess oxygen atoms within Bi-O layers, which provide holes into Cu-O planes 
nearby, will be lost to a high degree during the process of cleaving. 
However, if the cleaving is carried out at very low temperatures, the 
reduction of excess oxygen atoms, thus hole carriers, will be suppressed to 
some extent. After the approach of the STM tip to the cleaved surface, we 
obtained STM images of 512$\times $512 pixels in the constant height mode 
under various constant sample biases, $V_{s}$'s. The differential 
conductance $dI/dV$ was measured by using a standard lock-in technique with ac 
bias modulation of 3 mV and a frequency of 4 kHz.

\begin{figure}[htbp]
\begin{center}
\includegraphics*[width=0.70\linewidth,clip]{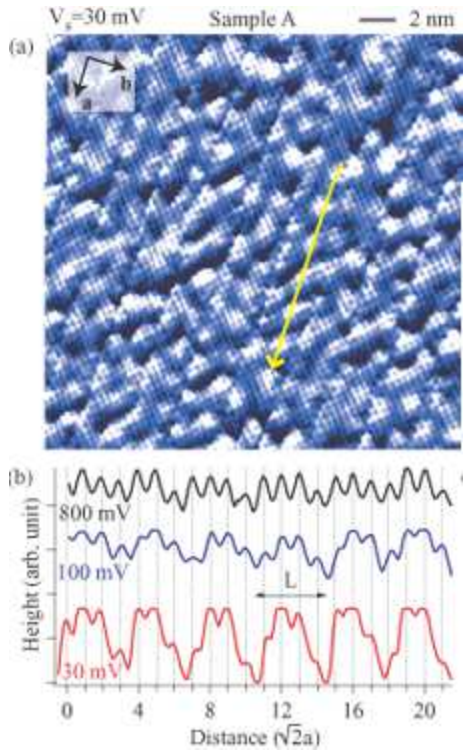}
\end{center}
\caption{(a) STM image of sample A at $T\sim $9 K, showing a 4a$\times $4a 
superstructure together with atoms and a 1D superstructure, inherent in the 
Bi-O plane, perpendicular to the b axis. The image was measured at a sample 
bias ($V_{s})$ of 30 mV and initial tunneling current ($I_{t})$ of 0.3 nA. (b) 
Profiles along the line at the same position in STM images at various bias 
voltages (yellow line in the top panel). The line was purposely taken to be 
perpendicular to axis b, that is, 45 degrees from the orientation of the 
4a$\times $4a superstructure so that the 1D superstructure of the Bi-O plane 
could not obscure the profile of the 4a$\times $4a superstructure. 
}
\label{f1}
\end{figure}

\begin{figure}[htbp]
\begin{center}
\includegraphics*[width=0.70\linewidth,clip]{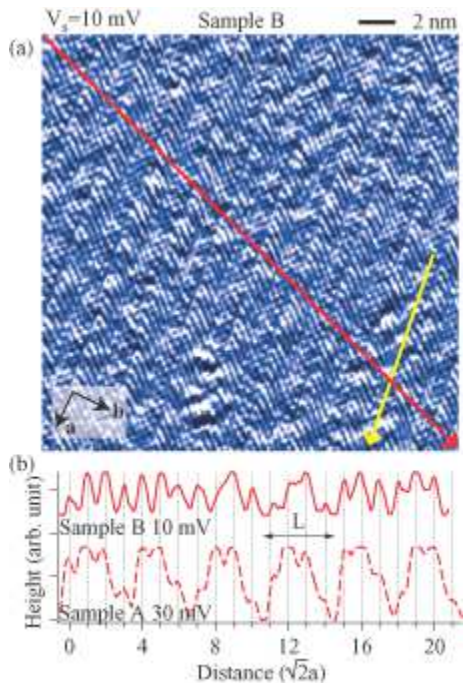}
\end{center}
\caption{(a) STM image of sample B at $T\sim $9 K, measured at $V_{s}$ = 10 mV, 
$I_{t}$ = 0.3 nA. (b) Line profile of the STM image at $V_{s}$=10 mV along the 
yellow line in the top panel. For comparison, the line profile of sample A 
at $V_{s}$=30 mV(Fig. 1(b)) is also shown (broken line). 
}
\label{f2}
\end{figure}

Shown in Fig. 1(a) is a typical STM image measured at $T\sim $9 K on sample A 
under a low-bias voltage ($V_{s}=30$ mV), which lies within the pairing gap 
as will be mentioned below. Low-bias STM imaging is crucial to investigating 
the electronic structure of the Cu-O plane within the pairing gap. Such 
imaging is especially favorable for the study of the Cu-O plane of Bi2212. 
This is because, when bias voltage $V_{s}$ lies within the semiconducting 
gap $E_{g }$($\sim $200 meV) of the Bi-O plane ($V_{s}<E_{g}$/e), electron 
tunneling occurs predominantly not between the STM tip and the Bi-O plane 
but between the STM tip and the Cu-O plane.$^{13)}$ Thus low bias STM images 
provide information about the Cu-O plane selectively. In Fig. 1(a), we can 
identify a Cu-O bond-oriented, 2D superstructure with a patched structure 
throughout the whole STM image, which was observed for both positive and 
negative biases. Fig. 1(b) shows line profiles of STM images under various 
$V_{s}$'s, taken along the yellow line shown in Fig. 1(a). (Note that the 
yellow line runs parallel to the 1D superstructure, which is inherent in the 
Bi-O plane, to avoid crossing it.) In Fig. 1(b), we can identify a 2D 
superstructure with a periodicity of 4a, namely a 4a$\times $4a 
superstructure, at $V_{s}$=30 and 100 mV in addition to the primitive 
lattice and its independence from the bias voltage. We also notice that the 
4a$\times $4a superstructure is more intense at lower bias voltage, and 
becomes very weak at a high $V_{s}$=800 mV where only a primitive lattice is 
observed. The low-bias STM image for sample B is shown in Fig. 2(a), and we 
can also identify the 2D superstructure, not throughout the cleaved surface, 
but over local regions on a nanometer scale. In Fig. 2(b), the line profile 
of the STM image with $V_{s}$=10 mV is shown for the region where the 2D 
superstructure appears most clearly. We can clearly see the superstructure 
with a periodicity of 4a, but its amplitude is evidently smaller than that 
observed for sample A. 

\begin{figure}[htbp]
\begin{center}
\includegraphics[width=0.75\linewidth,clip,clip]{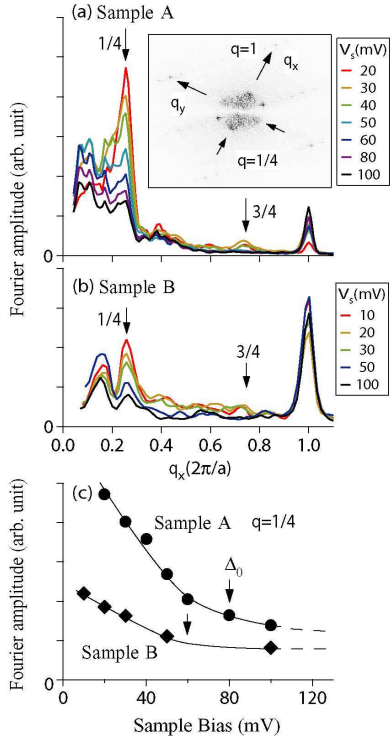}
\end{center}
\caption{(a) Line cuts of 2D Fourier maps of the STM image for sample A along 
(0, 0)-(0, $\pi )$ at various bias voltages. Inset is a 2D Fourier map of 
the real-space STM image (Fig. 1(a)). (b) Line cuts of the 2D Fourier map of 
the STM image for sample B (Fig. 2(a)). (c) Energy dependence of the Fourier 
amplitude at $q$ = 1/4. The arrows show the average sizes of the pairing gaps 
measured on the sample A and B. 
}
\label{f3}
\end{figure}

In Fig. 3, a 2D Fourier map $F(q_{x}, q_{y})$ of the real-space STM image 
taken at $V_{s}$=30 mV is shown for sample A. Besides Fourier peaks 
corresponding to the underlying primitive lattice (Bragg peaks) and the 1D 
superstructure of the Bi-O plane, two different Fourier peaks appear along 
both $q_{x}$ and $q_{y}$ axes: $\textbf{q}=(1/4, 0)2\pi /a$ and (3/4, 
0)2$\pi /a$, (0, 1/4)2$\pi /a$ and (0, 3/4)2$\pi /a$. The peaks along the 
$q_{x}$ axis are stronger than those along the q$_{y}$ axis. The Fourier 
peak at $q=3/4\times $(2$\pi $/a) indicates that the 4a$\times $4a 
superstructure has an internal structure with a period of 4a/3. For 
quantitative analysis of the Fourier maps, line cuts of the 2D Fourier map 
along the ($\pi $/a, 0) direction are shown as a function of $V_{s}$ (Fig. 
3(a)). The Fourier peaks at $q_{x}$= 1/4$\times $(2$\pi $/a) and 3/4$\times 
$(2$\pi $/a) are most intense at the lowest bias, 20mV, and decrease rapidly 
with the increase of $V_{s}$. Finally they tend to fade out above 
$V_{s}\sim $100 mV. It should be noted here that those Fourier peaks show 
no change in position and no broadening even if $V_{s}$ increases, 
indicating that the 4a$\times $4a superstructure observed in the present STM 
experiments is nondispersive. The correlation length determined from the 
full width at half-maximum of the Fourier peak is rather short, about 6 nm. 
The line cuts of 2D Fourier map are also shown for sample B along the ($\pi $/a, 0) 
direction (Fig. 3(b)). We can identify the 1/4$\times $(2$\pi $/a) 
peak up to 50 mV, while the 3/4$\times $(2$\pi $/a) peak is very weak. The 
intensity of the 1/4$\times $(2$\pi $/a) peak normalized with the Bragg peak 
is much weaker than that for sample A, consistent with the results of the 
line profiles of real-space STM images. The 1/4$\times $(2$\pi $/a) peak 
decreases with increased bias voltage and tends to fade out at bias voltages 
above $\sim $50 mV. Peak-like structures are observed at $q<0.2$ in A and B 
samples, but they show almost no energy dependence in peak intensity, 
meaning that these structures are irrelevant to the superstructure that we 
focused on (Fig. 3(a) and (b)). 

Shown in Fig. 5 is an STS spectrum, averaged over the distance of $\sim $40 
nm on the cleaved surface of sample A, which exhibits an intense, 
nondispersive 4a$\times $4a superstructure throughout the whole STM image. 
For comparison, a typical ZTPG spectrum reported for lightly doped Na-CCOC 
is also shown in Fig. 5.$^{2)}$ The STS spectrum of sample A has a structure 
very similar to the ZTPG; that is, an asymmetric V-shaped gap without 
coherence peaks. Width $\Delta _{0}$ of the present ZTPG is $\sim $80 meV 
on the positive bias side. In Fig. 6, the spatial dependence of STS spectra 
is shown for the cleaved surface of sample B whose STM image exhibits the 
4a$\times $4a superstructure locally. It should be stressed that the STS 
spectra of sample B are very homogeneous and show a gap structure of the 
d-wave type. The gap width, $\Delta _{0}$, is $\sim $60 meV, though it 
tends to be slightly enhanced over the regions exhibiting the 4a$\times $4a 
superstructure clearly. The specific, homogeneous d-wave gap means that the 
doping level is rather homogeneous in sample B and the hole pairs are 
uniformly formed all over the cleaved surface. However the electronic 
4a$\times $4a superstructure appears only locally, as mentioned above. The 
dynamical 4a$\times $4a charge order probably evolves throughout the cleaved 
surface of sample B, and some kind of disorder will pin down the dynamical 
charge order locally. It is worthwhile to point out here that in both 
samples A and B, the value of $\Delta _{0}$ roughly corresponds to the STM 
bias voltage below which the 4a$\times $4a superstructure can clearly be 
observed in STM imaging. This fact implies the possibility that in-gap 
states, namely the hole pairs, will contribute to the formation of the 
4a$\times $4a superstructure. In high-$T_{c}$ cuprates, the gap size 
increases with the decrease in the doping level.$^{14, 15)}$ Thus, the 
larger gap size measured on the cleaved surface for sample A than for sample 
B means that the doping level of the cleaved surface of sample A is lower 
than that of sample B, though both samples are cut from the same single 
crystal. The 4a$\times $4a superstructure is very intense in sample A and 
appears throughout the cleaved surface, whereas it is relatively weak in 
sample B and appears locally over the cleaved surface, as mentioned above. 
Furthermore, the low-bias STM image previously reported for nearly optimally 
doped Bi2212 with $\Delta _{0}\sim $35 meV exhibited no 4a$\times $4a 
superstructure.$^{13)}$ These results indicate that a low hole doping level 
favors the formation of the 4a$\times $4a superstructure. It should be noted 
here that the cleaved surface of sample A has more missing atom rows within 
the cleaved Bi-O plane than that of sample B, which can be clearly seen in 
the high-bias STM images, predominantly reflecting the electronic structure 
of the Bi-O plane. (The high-bias STM images of the present samples will be 
published elsewhere.) If the missing atom rows are introduced within the 
Bi-O plane during the cleaving process, excess oxygen atoms will also be 
removed around the missing atom rows at the same time, which leads to the 
reduction of the average hole doping level of the Cu-O plane. Thus the high 
density of the random missing atom rows in the cleaved surface of sample A 
may be a possible reason why the average hole doping level is lower in the 
cleaved surface of sample A than for sample B, though both samples were cut 
from the same single crystal. The low density of the random missing atom 
rows in the cleaved surface of sample B can also explain its homogeneous 
electronic structure, namely the homogeneous paring gap.

\begin{figure}[htbp]
\begin{center}
\includegraphics*[angle=-90, width=0.7\linewidth,clip]{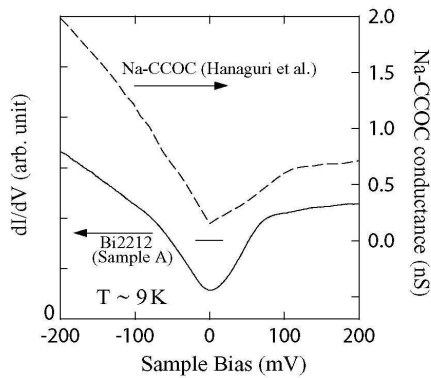}
\end{center}
\caption{STS spectrum averaged over the distance of 40 nm on the cleaved 
surface of sample A at $T\sim $9 K. A typical ZTPG spectrum reported for 
Na-CCOC (x=0.12) is also shown for comparison (broken line).$^{2)}$
}
\label{f4}
\end{figure}

\begin{figure}[htbp]
\begin{center}
\includegraphics*[width=0.7\linewidth,clip]{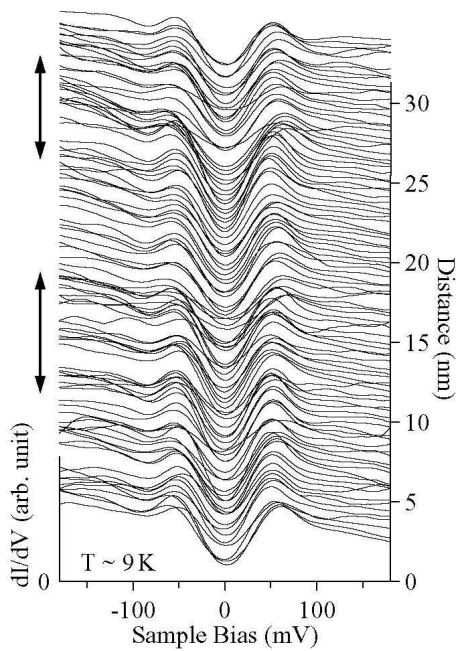}
\end{center}
\caption{Spatial dependence of STS spectra for sample B, taken at $T\sim $9 K along 
the red line in Fig. 2(a). Two-headed arrows beside the spectra indicate the 
regions where the 4a$\times $4a superstructure is clearly observed.
}
\label{f5}
\end{figure}

In the present experiments of low-bias STM imaging, namely Cu-O plane 
selective imaging, on underdoped Bi2212,$^{13)}$ we confirmed that an 
intense, nondispersive 4a$\times $4a superstructure was formed within the 
Cu-O plane. The 4a$\times $4a superstructure has a substructure with a 
periodicity of 4a/3 and is clearly observed within the pairing energy gap. 
Such features of the present 4a$\times $4a superstructure are in good 
agreement with those of the 4a$\times $4a electronic charge order reported 
for lightly doped Na-CCOC.$^{2)}$ The intense, nondispersive 4a$\times $4a 
electronic charge order is different in origin from the dispersive 2D 
modulation caused by quasiparticle scattering interference.$^{5)}$ The 
present nondispersive electronic charge order is consistent with the 
findings of Howald et al. in the LDOS maps for the SC state of Bi2212.$^{7, 8)}$ 
They claim that the nondispersive superstructure results from the 
formation of a static stripe order. The present results can be explained in 
terms of a static stripe order except for the experimental finding for 
sample B that the paring gap is influenced only slightly even over the 
region where the dynamical 4a$\times $4a charge order is locally pinned  
down. Since the formation of the static stripe order causes the 
superconductivity to seriously deteriorate, as demonstrated for 
La$_{2 - x}$Ba$_{x}$CuO$_{4}$ and La$_{1.6 - x}$Nd$_{0.4}$Sr$_{x}$CuO$_{4}$ systems 
where simple 1D strips develop, the paring gap would be seriously influenced 
over the region where the 4a$\times $4a electronic charge order is pinned 
down. The present 4a$\times $4a electronic charge order, clearly appearing 
within the pairing gap, is not inconsistent with models for the electronic 
charge order due to pair density waves, electronic supersolids, or 
paired-hole Wigner crystallization.$^{16 - 22)}$ 

This work was supported in part by Grants-in-Aid for Scientific Research and 
the 21$^{st}$ century COE program ``Topological Science and Technology'' 
from the Ministry of Education, Culture, Sports, Science and Technology of 
Japan.


\begin{thebibliography}{99} 
\bibitem{1} M. Vershinin, S. Misra,S. Ono, Y. Abe, Y. Ando and A. Yatsudani: Science \textbf{33} (2004) 1995.
\bibitem{2} T. Hanagri, C. Luplen, Y. Kohsaka, D-H. Lee, M. Azuma, M. Takano, H. Takagi and J. C. Davis: Nature \textbf{430} (2004) 1001.
\bibitem{3} K. McElroy, R. D-H. Lee, J. E. Hoffmann, K. M. Lang, J. Lee, E. W. Hudson, H. Eisaki, S. Uchida and J. C. Davis: cond-matt/0406491.
\bibitem{4} J. E. Hoffman, E. W. Hudson, K. M. Lang, V. Madhavan, H. Eisaki, S. Uchida, J. C. Davis: Science \textbf{295} (2002) 466.
\bibitem{5} J. E. Hoffmann, K. McElroy, D-H. Lee, K. M. Lang, H. Eisaki, S. Uchida and J. C. Davis: Science \textbf{297} (2002) 1148.
\bibitem{6} K. McElroy, R. W. Simmonds, J. E. Hoffmann, D-H. Lee, K. J. Orenstein, H. Eisaki, S. Uchida and J. C. Davis: Nature \textbf{422} (2003) 592. 
\bibitem{7} C. Howald, H. Eisaki, N. Kaneko, M. Greven and A. Kapitulnik: Phys. Rev. B \textbf{67} (2003) 014533.
\bibitem{8} A. Fang, C. Howald, H. Eisaki, N. Kaneko, M. Greven and A. Kapitulnik: Phys. Rev. B \textbf{70} (2004) 214514.
\bibitem{9} S. A. Kivelson, E. Fradkin, and V. J. Emery: Nature \textbf{393} (1998) 550.
\bibitem{10} M. Bosch, W. van Saarloos, and J. Zaanen: Phys. Rev. B \textbf{63} (2001) 092501.
\bibitem{11} M. Oda, H. Matsuki, and M. Ido: Solid State Commun. \textbf{74} (1990) 1321.
\bibitem{12} T. Nakano, M. Oda, C. Manabe, N. Momono, Y. Miura, M. Ido: Phys. Rev. B \textbf{49} (1994) 16000.
\bibitem{13} M. Oda, C. Manabe and M. Ido: Phys. Rev. B \textbf{53} (1996) 2253.
\bibitem{14} M. Oda, K. Hoya, R. Kubota, C. Manabe, N. Momono, T. Nakano, M. Ido: Physica C \textbf{281} (1997) 135.
\bibitem{15} N. Miyakawa, P. Guptasarma, J. F. Zasadzinski, D. G. Hinks, K. E. Gray: Phys. Rev. Lett. \textbf{80} (1998) 157.
\bibitem{16} H. C. Fu, J. C. Davis and D-H Lee: con-matt/0403001.
\bibitem{17} M. Vojta: Phys. Rev. B \textbf{66} (2002) 104505.
\bibitem{18} H-D Chen, O. Vafek, A Yazdani and S-H Zhang: Phys. Rev. Lett. \textbf{93} (2004) 187002.
\bibitem{19} S. Sachedev and E. Demler: Phys. Rev. B \textbf{69} (2004) 144504. 
\bibitem{20} Z. Tesanovic: Phys. Rev. Lett. \textbf{93} (2004) 217004.
\bibitem{21} M. Franz: cond-mat/0409431.
\bibitem{22} P. W. Anderson: con-matt/0406038.
\end{thebibliography}
\end{document}